\documentclass{aa} 
\usepackage{graphicx}
\begin{document}
   \title{The color signature of the transit of HD 209458: Discrepancies
   between stellar atmospheric models and observations}

   \author{B. Tingley \inst{1,2}, C. Thurl \inst{1,2}, and P. Sackett
   \inst{1,2}}

   \offprints{B. Tingley}

   \institute{\inst{1} Research School of Astronomy and Astrophysics, ANU
               Cotter Road, Weston, Canberra ACT 2611, Australia \\
             \inst{2} ANU Planetary Science Institute c/o Mount Stromlo Observatory,
               Cotter Road, Weston ACT 2611, Australia \\ 
              \email{tingley@mso.anu.edu.au, cthurl@mso.anu.edu.au, Penny.Sackett@anu.edu.au}}

   \date{Received X, 2005; accepted Y, 200Z}

   \abstract{Exoplanetary transits produce a double-horned color 
   signature that is distinct from both binaries and blends and
   can thus be used to separate exoplanets from false positives in transit
   searches. Color photometry with precision sufficient to detect this signal
   in transits of HD 209458 is available in the literature. Analysis of these
   observations reveals that, while the signature does exhibit the expected
   shape, it is significantly stronger than PHOENIX atmospheric models predict.

   }
				       
   \maketitle
%

\section{Introduction}
The radial velocity technique is most commonly used technique for the
classification of transiting exoplanet candidates. It measures the velocity
shift of the parent star, which allows estimates the mass of the transiting
companion via observations of a radial velocity shift in the lines of the host
star. However, this technique is not ideal. Not only is it time- and
resource-consuming, but it fails to classify some candidates. These
unclassifiable candidates can be either too faint to be observed with the
precision necessary to identify the signal, too active to allow the signal to
be identified over the stellar noise, or are actually blends. In many cases,
blends leave no spectral fingerprint and thus cannot be discerned from the
other phenomena and represent the greatest obstacle to transit surveys with
the potential to discover terrestrial planets.

Other techniques exist to make this classification. The Rossiter-McLaughlin
effect manifests itself as a perturbation of the radial velocity of the parent
star during a transit, allowing the transiting body to be classified (Worek
(\cite{worek}). However, it does suffer from most of the same problems as the
radial velocity technique. Seager and Mall\'{e}n-Ornelas (\cite{seager})
describe a technique that analyzes the shape of the transit to determine the
mass of the transiting body, for which is blends produce anomolous
results. This technique requires high-precision photometry and utilizes
assumptions about the mass-radius relationship for the lower main sequence,
which do not necessarily hold in all cases. Torres et al. (\cite{torres})
propose a technique involving detailed modelling of the full light curve to
test blend scenarios, which helped in identifying OGLE-TR-33 as a blend.

The color change exhibited by the central star during transit can also be used
to make this classification, as exoplanets, grazing binaries and blends all
have own, distinct color signature. First realized by Rosenblatt
(\cite{rosenblatt}) and further developed by Borucki \& Summers
(\cite{borsum}) and discussed in connection with hot Jupiters by Sackett
(\cite{sackett}), a non-grazing exoplanetary transit will exhibit a sharply
peaked double-horned color profile, which models predict will have an
amplitude on the order of 10\% of the transit depth.  This is markedly
different from the broad, single-peaked profiles of binary stars and blends
(Tingley \cite{tingley}). The strengths and shapes of these signatures are
highly dependent on various factors: the differential limb darkening between
wavebands, orbital characteristics, the relative sizes of the transiting
objects and the color differences between the eclipsing (normally a background
eclipsing binary star) and constant (non-eclipsing) components.

Given that typical giant exoplanet transits have a depth of 1-2\% percent,
exoplanetary color signatures should be on the order of a few millimags. At
present, good ground-based photometry can have a precision better than 1
millimag per observation (see, for example, Jha et al.  \cite{jha}). This
means that these signatures should be observable from the ground and should
therefore be present in high precision multi-color observations of the transit
of HD 209458 already available in the literature. The photometric
possibilities from space are even more promising. Precisions approaching 0.1
mmag are possible (Brown et al. \cite{brown}) with existing instruments, which
can help to verify the strength of the signature. In the absence of suitable
ground-based data, this can be used to establish if the signature is
detectable with ground-based photometry.

\section{Photometry of the transit of HD 209458}

Three different sets of observations of HD 209458 are available in the
literature that have sufficient precision to be useful for this study: the HST
observations of Brown et al. (\cite{brown}), the simultaneous Stromgren
photometry of Deeg et al. (\cite{deeg}), and the $BVIRZ$ observations of Jha
et al. (\cite {jha}).

\subsection{HST observations}

In 2000, the HST was utilized in an unprecedented fashion in order to obtain
extremely high precision photometry of the HD 209458 transit. By using the
Space Telescope Imaging Spectrograph (STIS) on a small, relatively featureless
portion of the spectrum (from 5813 to 6382 Angstroms), Brown et
al. (\cite{brown}) were able to obtain a photometric precision of $1.1 \times
10^{-4}$ per sample, with one sample every $80s$, for the transit of HD
209458.  This surpasses the capabilities of any ground-based instrument by a
factor of approximately ten. However, the observations were taken only in this
single passband, and so alone cannot be used to identify a color change during
the transit. The authors did split their spectrum into ``red'' (the reddest
300 Angstroms) and ``blue'' (the bluest 300 Angstroms) and noted that a feature
was present in the ``red''-''blue'' color.  The weakness of the signature is
directly attributable to the fact that there is little wavelength
separation in this ``color''. However, when combined with ground-based data,
the extremely high precision and temporal coverage of the HST data
allows the color signatures to be assessed and analyzed.

\subsection{Observatorio Sierra Nevada observations}

Another useful data set was created by Deeg et al. (\cite{deeg}), which
contains simultaneous Stromgren $u$, $v$, $b$ and $y$ photometry using the
0.9m telescope at the Observatorio Sierra Nevada in Spain. After
sigma-clipping, this data set includes 129 in-transit plus 29 out-of-transit
observations in all bands. The precision of these data was about 4 mmag in $u$
and $y$ and 2 mmag in $v$ and $b$, as measured from the out-of-transit
scatter. While the data are of excellent quality, the two filters with the
highest precision do not have a large separation in wavelength, with only 600
Angstroms between $v$ and $b$. Moreover, there appears to be some systematic
noise present, which is most prominent in $u$ and $y$, less so in $b$ and
almost non-existent in $v$. Readily visible in Figure 1 of Deeg et
al. (\cite{deeg}), these variations are sufficiently strong to obscure the
expected exoplanetary signature in all but $v$.

\begin{figure*}
   \centering
   \includegraphics[width=0.5\textwidth]{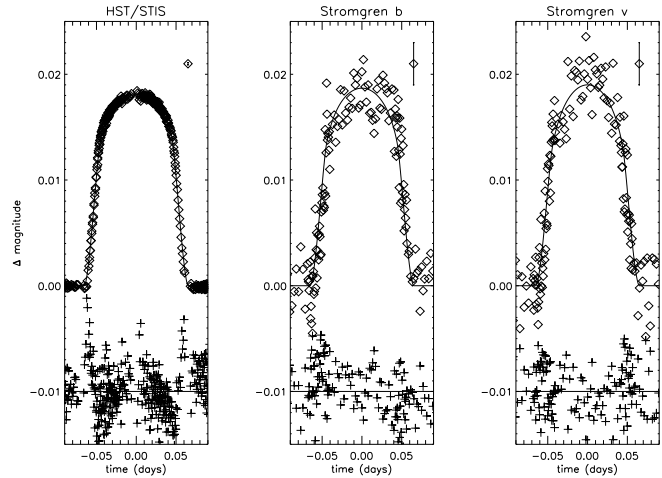}
   \caption{Data (diamonds), the modeled fits (solid
lines), residuals (crosses) are shown for the transit of HD 209458.
Typical error bars are shown in the
upper right hand corner of each plot. The model transits were determined
by using the best fit of 362 models to the HST data. Note that there is still
a small amount of symmetric residual in the HST fit, which has been
increased by a factor of 10 for clarity. Note also the strongly
asymetric residual in the $b$ passband, the cause of which is unknown.}
\end{figure*}

\subsection{Hawai'i 2.2 m and Hawaii 0.6 m observations}

Useful observations from the University of Hawai'i 2.2m and 0.6m were
published by Jha et al. (\cite{jha}). They include $B$, $V$, $I$, $R$ and $Z$
photometry of a transit of HD 209408 in 1999, with precisions of 0.8 mmag, 1.6
mmag, 1.2 mmag and 0.8 mmag respectively. The event was observed 43 times in
$B$ from the Hawaii 0.6 m and 17 times in each of the other filters from the
Hawaii 2.2 m. This data set had the potential to be the most useful, as it is
comprised of simultaneous, high-precision photometry through multiple filters
with broad separation in wavelength. The purpose of the Jha et al project,
however, was to observe slight differences in transit depth in different
bands, in order to get a better estimate of the radius of HD 209458b. As such,
the data do not capture the ingress of the transit and additionally have poor
time resolution.  These factors mean that these data unsuitable for the
identification of a color signature that occurs over relatively short time
scales, especially considering that the observations alternated filters and
switched to comparison stars. The lack of an ingress makes it hard to compare
these data with other data sets, as systematic errors can easily produce a
single horn in color typical of those caused by an exoplanet during
ingress/egress. The presence of a second horn at the proper time is a
necessary verification.

\section{Modeling}

The atmosphere of HD 209458 was modeled using state-of-the-art PHOENIX grid
models (v. 2.6) (Hauschildt et al. \cite{phoenixI}, Hauschildt, Allard \&
Baron, \cite{phoenixII}, Hauschildt \& Baron \cite{baron}) obtained through an
ongoing collaboration with Peter H. Hauschildt. PHOENIX models show
significantly different limb darkening behavior than that described by
analytic limb darkening laws (Bryce, Hendry \& Valls-Gabaud \cite{bryce};
Claret \cite{claret03}). The models used to fit the transit are
one-dimensional, assume Local Thermodynamic Equilibrium (LTE) and use
spherically-symmetric radiative transfer and dynamic opacity sampling. The
intensity profile is calculated with a wavelength resolution of $\le$ 1\AA ~at
99 angular points. These points are distributed evenly in $\cos\theta$ for
most of the inner parts of the stellar disk, where $\theta$ is the emergent
angle. However, angular sampling increases toward the limb, where greater
changes in the intensity profile occur.

The model atmospheres were generated for several different effective
temperatures (5900$K$, 6000$K$, 6100$K$), surface gravities ($\log g = 4.0,
4.5$) and metallicities (-0.5, 0.0, +0.5). The model atmospheres were
converted to limb darkening profiles by convolving the stellar intensity
profiles with the appropriate filter passbands. The limb-darkening profiles
were then used to create modeled transit shapes, leaving the exoplanet-star
radius ratio $(\frac{R_p}{R_\star})$, the impact parameter $(i_p)$ (the
distance from center of the star to the point of nearest approach of the
exoplanet's projected path across the disk of the star, in units of stellar
radii) and the duration of the transit as free parameters. The code that
modeled the transits was designed to include subpixels to minimize the effects
of pixellation on the limbs of the exoplanet and star. This may otherwise
cause small but noticeable errors. A simple $\chi^2$ minimalization was used
to find the best-fit model to the HST data. The observed properties of HD
209458 are uncertain: ($[{\rm Fe/H}]=0.00\pm0.02$, $M_* = 1.1\pm0.1 M_\odot$
and $R_* = 1.2\pm0.1 R_\odot$, which yield $\log g = 4.36 \pm 0.05$, $T_{\rm
eff} = 6000 {\rm K} \pm 50 {\rm K}$ (Mazeh et al. \cite{mazeh}). Other groups
have found slightly different values: Cody \& Sasselov (\cite{cody}) reported
$M_* = 1.06\pm0.11 M_\odot$ and $R_* = 1.18\pm0.12 R_\odot$, which yield $\log
g = 4.32 \pm 0.09)$, Fischer \& Valenti (\cite{fischer}) reported $M_* = 1.14
M_\odot$ and $R_* = 1.12 R_\odot$ with an error in $\log g$ of 0.06 dex, which
yield $\log g = 4.40 \pm 0.06$ and Valenti \& Fischer (\cite{valenti})
reported $[{\rm Fe/H}]=+0.02\pm0.03$. These values for $\log g$ and $[{\rm
Fe/H}]$ are all within nearly one sigma of one another and futhermore
differences in these parameters at this level have a very small effect on
transit shape -- changing metallicity or temperature by 0.5 dex results in a
change in a transit shape of less than 1\% of the transit depth, while a
similar change in surface gravity changes the shape by less than
0.1\%. Therefore, it is reasonable to use intermediate values of temperature
(5950$K$, 6000$K$, 6050$K$) and surface gravity ($\log g = 4.30, 4.36, 4.42$)
to reflect these measurements and their uncertainties. These were then
interpolated from the original models. The best-fit system parameters, using
only the HST data as they are far more stable and precise than the Deeg data,
were ($T_{\rm eff} = 6050$ K, $[{\rm M}/{\rm H}] = -0.5$, $\log g = 4.42$,
$R_p = 0.122 R_\star = 1.36 R_{\rm J}$, $i = 86.1^\circ$).  The result of this
analysis can be seen in Figures 1 (fits to transit shape) and 2 (resulting
modeled color change).

Since the LTE PHOENIX models produced a color signature that accurately
described the shape but not the amplitude of the observed signature, we
implemented a fully non-LTE PHOENIX model in order to study if 
this could explain the observed discrepancies. As the non-LTE models have an
immense computational load, only one was created and compared to the LTE model
with the same parameters. As can be seen in Figure 3, a fully non-LTE model
does exhibit a stronger color signature than an LTE model (solar metallicity,
$T_{eff} = 6000$ K, and ${\rm log} g = 4.50$, with planetary parameters $R_p
= 0.122 R_\star = 1.36 R_{\rm J}$ and $i = 86.1^\circ$), but the difference is
marginal compared to the differences between model and observations.

\section{Results}

The data from Deeg et al. (\cite{deeg}) alone do not reveal the exoplanetary
color signature. The $u$ and $y$ observations are too noisy, especially at the
ingress and egress of the transit, which are most critical for this
analysis. The $b$ observations contain an asymmetric variation that is likely
due to undetermined systematic effects. Only the $v$ is sufficiently precise
and stable.  Moreover, even without the systematics in $b$, the $b$ and $v$
passbands (centered at 4100 \AA ~and 4700 \AA ~respectively) do not have
enough chromatic separation to produce a strong signature.  The expected
signal in $b-v$ from the models is less than 1 mmag from peak to trough. In
principle, this is detectable with these data, given 145 observations with a
precision in this color of approximately 3 mmag. However, the systematic noise
obscures the faint expected signature.

The HST data must be combined with the Deeg data, therefore, to reveal
the explanetary color signature.
Figure 2 shows modeled and observed $b-HST$ and $v-HST$. The $b-HST$
shows a similar trend as observed in the residual for $b$ in Figure 1. The
$v-HST$ plot clearly shows a color change that has a shape consistent with that
predicted by the atmospheric models, but with a stronger amplitude.
There was also some small amount of residual between the HST data and
the model, visible in Figure 1.

\begin{figure}
   \centering
   \includegraphics[width=0.5\textwidth]{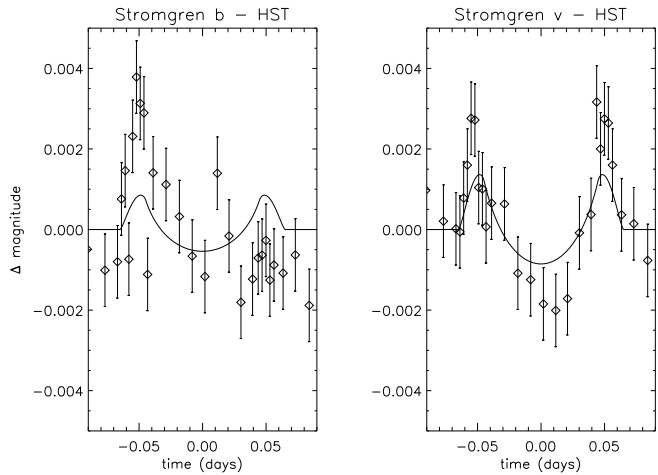}
   \caption{This figure shows the $b-HST$ and $v-HST$ color change during the
   transit, along with the signatures from LTE PHOENIX models. Note that the
   systematic noise evident in $b$ is also evident in the $b-HST$ color change.
   The observed $v-HST$ color change is very consistent in shape, if not in
   amplitude, to the model.}
\end{figure}

\begin{figure}
   \centering
   \includegraphics[width=0.5\textwidth]{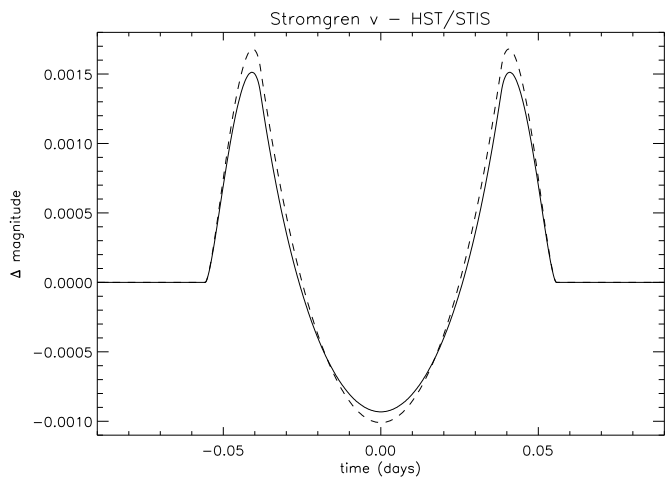}
   \caption{This figure shows the $v-HST$ color change for a star with solar
   metallicity, $T_{eff} = 6000$ K, and ${\rm log} g = 4.50$, with planetary
   parameters $R_p = 0.122 R_\star = 1.36 R_{\rm J}$ and $i = 86.1^\circ$. The
   solid line shows a model assuming LTE while the dashed line is a fully
   non-LTE model. Note that the latter demonstrates a stronger signature, yet
   still predicts a significantly weaker signature than the one observed.}
\end{figure}

\section{Conclusion}

An exoplanetary color signature is clearly visible in the transit of HD 209458
in $v-HST$.  Unfortunately, the data from Deeg et al. (\cite{deeg}) alone did
not exhibit an exoplanetary signature, as the data were either too noisy or
not separated enough in wavelength from the other passband. Even so, the
signature that was detected using the HST data and the best of the
ground-based data is strong enough that it could be observed from ground, in
the absence of these systematics and in more widely separated passbands.

The color signature detected in the transit data of HD 209458 in our analysis
was consistent in shape, but larger in amplitude than that expected using
either LTE or fully non-LTE PHOENIX models, though the non-LTE models were
superior.  The detected signature was considerably stronger in $v - {\rm
HST/STIS}$, with an amplitude of approximately 5 mmag observed against 2.4
mmag predicted by the LTE model and 2.7 mmag predicted by the non-LTE model.
This increases the viability of the color signature as a technique for
classifying transit candidates, as the observed signature exceeds any
prediction, being closer to 30\% of the transit depth than 10\%, the generally
quoted valued.

The system parameters we derived are essentially identical to those found by
Brown et al. (\cite{brown}), despite the fact that they used a standard
stellar limb darkening law, fitting its coefficients as free paremeters.  Thus
the use of PHOENIX models does not adversely affect the estimation of transit
parameters.  The discrepancy in the amplitude of the color signature between
the HD 209458 and PHOENIX models, suggests, if confirmed in other systems,
that an important component to the factors that influence color near the
stellar limb may still be absent in these models.

\begin{acknowledgements}
The authors would like to thank Peter Hauschildt, who provided
his assistance and considerable expertise in the setting up of the
PHOENIX models.
\end{acknowledgements}

\end{document}